\documentclass{eptcs}

\usepackage[latin1]{inputenc}
\usepackage{bussproofs}
\usepackage{amssymb}
\usepackage{graphicx}
\usepackage{latexsym}
\usepackage{amssymb}
\usepackage{amsmath}
\usepackage{qtree}
\usepackage{url}
\usepackage{cite}
\usepackage{breakurl}

\providecommand{\firstpage}{1}

\makeatletter
\def \equalsfill{$\m@th \mathord=\mkern-7mu
\cleaders \hbox{$\!\mathord=\!$}\hfill \mkern-7mu \mathord=$}
\makeatother

\title{Automatic Generation of Proof Tactics for~Finite-Valued~Logics
\thanks{The author acknowledges financial support by CNPq and CAPES.
        Dalmo Mendon\c{c}a and Lucas de Lima,
        both also financially supported by CNPq,
        wrote substantial chunks of the implementation code
        and carefully proof-read the paper, at different stages.} }
\author{Jo{\~a}o Marcos
\institute{LoLITA and DIMAp / UFRN, Brazil, and\\
           Institut f\"ur Computersprachen (E1852), TU-Wien, Austria}
\email{jmarcos@dimap.ufrn.br} }
\def\titlerunning{Automatic Generation of Proof Tactics for~Finite-Valued~Logics}
\def\authorrunning{Jo{\~a}o Marcos}

\begin{document}
\maketitle

\renewcommand{\arraystretch}{1.2}

\begin{abstract}
\noindent A number of flexible tactic-based logical frameworks are
nowadays available that can implement a wide range of
mathematical theories using a common higher-order metalanguage.
Used as proof assistants, one of the advantages of such powerful
systems resides in their responsiveness to extensibility of their
reasoning capabilities, being designed over rule-based
programming languages that allow the user to build her own
`programs to construct proofs' --- the so-called proof tactics.

The present contribution discusses the implementation of an
algorithm that generates sound and complete tableau systems for a
very inclusive class of sufficiently expressive finite-valued
propositional logics, and then illustrates some of the challenges
and difficulties related to the algorithmic formation of automated
theorem proving tactics for such logics. The procedure on whose
implementation we will report is based on a generalized notion of
analyticity of proof systems that is intended to guarantee
termination of the corresponding automated tactics on what
concerns theoremhood in our targeted logics.\smallskip

\end{abstract}

\section{Introduction}

The early history of the LCF family of theorem provers, first
implemented as proof checkers by Robin Milner in the early 70s,
based on Dana Scott's Logic for Computable Functions, can be said
to be essentially an evolution of Alonzo Church's original proposal of a simple
theory of types, developed three decades before
(cf.~\cite{pau:camLCF}). Arguably, though, their great success as
generic logical frameworks for the specification of a wide range
of useful mathematical theories within a unified setting came in
fact from later developments, namely: (1)~the design of an
accompanying powerful type-safe functional language that would
allow for the needs of the theorem-proving community to be quite
naturally expressed; (2)~the decision to use a constructive
higher-order logic as the underlying metalanguage and to use
higher-order unification as the underlying mechanism in which to
specify diverse genera of inference systems as theories written
in a common framework.  The programming language that was
designed in that process, \texttt{ML}, was intended to give
support to the expression of higher-order abstract syntax
for the definition and manipulation of object-logics, as well as to
advanced pattern-matching capabilities for the definition and
manipulation of abstract high-level datatypes. From the point of
view of theorem-proving, such flexible datatypes were to allow
for the representation of useful objects such as
\textit{formulas}, \textit{theorems\/} or even \textit{proofs},
as well as some strategical operations over those objects, called
\textit{tactics}, that represented subgoaling strategies used in
the construction of proofs. Higher-order operations for combining
tactics and taking stricter control of the result of proof-search
procedures were also to be made available as the so-called
\textit{tacticals}.  A modern heir of the LCF-style family of
proof assistants and tactical provers, allowing for both
interactive and automated reasoning, is the system
\texttt{Isabelle} (cf.~\cite{Nipkow-Paulson-Wenzel:2002}), which
will be utilized in what follows.

A simple and elegant deductive formalism for the specification of
proof procedures for both classical and non-classical logics is
provided by the refutation-oriented method of \textit{tableaux\/}
(cf.~\cite{smu:FOL}). In the classical bivalent propositional
case, the inference rules of (signed or unsigned) tableau systems
are based on adequate versions of a \textit{subformula
principle\/} that guarantees that the overall complexity of the
involved formulas decreases as tableau rules are applied in the
construction of a tableau derivation. The resulting collection of
rules, in that case, is said to be \textit{analytic}, and
decidability, in general, follows from that.  Indeed, analytical
proof procedures eliminate in particular the use of the so-called `cut rule'
(which often presupposes some ingenuity from the proof designer)
and are very useful for
automation as they greatly facilitate the finding of proofs.  On
the other hand, exactly because they eliminate cut, such
procedures render the expression of proof lemmas more difficult,
if not outright impossible. However, this limitation can often be
negotiated with an additional gain in the speed-up of the
corresponding derivations if one considers systems allowing for
the so-called `analytic cuts' (cf.~\cite{dag:mon:taming}).  In
one way or another, the objective is to define a rule-based
framework for propositional logics in which the termination, with
more or less efficiency, of a given theorem-proving task is
guaranteed at the outset.

In \cite{cal:car:con:mar:humbug:05:full} an algorithm was devised
to extract bivalent (in general, non-truth-functional)
characterizations for an extensive class of finite-valued
propositional logics and then turn those characterizations into
classic-like adequate tableau systems for those logics.  By a
`bivalent' characterization of a logic, here and in that paper, we mean a
collection of interpretation mappings that takes only
\textit{two\/} `logical' values into consideration, in spite of
the many `algebraic' values that might be used by the logic's
original multi-valued truth-functional semantics --- the role of
the extraction algorithm is to guarantee that both the bivalent
and the finite-valued characterization end up determining the same
entailment relation.  We have used~\texttt{ML} to implement the
mentioned algorithm in~\cite{mar:men:tfaae4fvl},\footnote{Check
also \url{http://tinyurl.com/5cakro}.} and the output of our
program is an \texttt{Isabelle} theory which can be used for
computer-assisted proofs of theorems and derived rules of the
corresponding finite-valued logics.  Such proof systems,
automatically extracted from the sets of truth-tables taken as
input by our program, contained a non-eliminable version of the
cut rule, and in fact no detailed proof was presented then that
analytic cuts, for instance, would suffice for every proof system generated by
the above mentioned algorithm. An improved axiom extraction
algorithm has recently been proposed in~\cite{ccal:mar:09a}, though, for
the same class of logics, in which cut \textit{is\/} eliminable.
The latter algorithm has some remarkable features, being based on
non-standard complexity measures that are intended to guarantee the
analyticity of its output, once one uses such measures to formulate
convenient proof strategies.  The paper~\cite{ccal:mar:09b} shows in detail how
that same axiom extraction mechanism can be extended for
\textit{any\/} finite-valued logic, irrespective of the
expressiveness of its original language.  The present paper
employs an illustration of this procedure to briefly report on the
challenges and difficulties related to the implementation of the
mentioned novel algorithms, having again as output
\texttt{Isabelle} theories, but this time enhanced with the
automatic formation of cut-free proof tactics for the complete automation
of the corresponding theorem-proving tasks.

\section{Tableaux}

A tableau system is both a proof and a counter-model building
procedure based on the construction of refutation trees. A
tableau rule is a schematic tree modifier, and its application
allows us, given a branch in which we find instances of the
rule's heads, to extend the leaf of this branch by considering
all the possibilities provided by the corresponding instances of
the rules's daughters.
For an example, the classical tableau rules
for negation and implication can be
represented as:
{\small
\begin{equation}\label{ClR}
\Tree
    [.{ $F{:}(\neg\alpha)$ }
      [.{ $T{:}\alpha$ }
      ]
    ]
\hspace{1.5cm}%
\Tree
    [.{ $T{:}(\neg\alpha)$ }
      [.{ $F{:}\alpha$ }
      ]
    ]
\hspace{1.5cm}%
\Tree
    [.{ $F{:}(\alpha\to\beta)$ }
      [.{ $T{:}\alpha $\\$ F{:}\beta$ }
      ]
    ]
\hspace{1.5cm}%
\Tree
    [.{ $T{:}(\alpha\to\beta)$ }
      [.{ $F{:}\alpha\hspace{.1cm} \ $}
      ]
      [.{ $T{:}\beta$}
      ]
    ]
\end{equation}%
}%
\noindent This means, for instance, that a branch containing a
signed formula of the form $F{:}(\alpha\to\beta)$ may be extended
by adding in sequence new nodes of the form~$T{:}\alpha$
and~$F{:}\beta$. Similarly, a branch containing a signed formula
of the form $T{:}(\alpha\to\beta)$ may be extended in two
different ways, both by adding a new node of the form~$F{:}\alpha$
and by adding a new node of the form $T{:}\beta$. The semantic
reading of such rules is obvious. The following \textit{closure
rule}, syntactically expressing an unobtainable semantic situation, completes
the characterization of classical logic:
{\small
\begin{equation}\label{ClRl1}
\Tree
    [.{ $T{:}\alpha$ \\ $F{:}\alpha$}
      [.{ $*$}
      ]
    ]
\end{equation}
}
The rule is intended to say that a branch that contains an
occurrence of the formula~$\alpha$ labelled with the sign~$T$ and
an occurrence of the same formula labelled with the sign~$F$ may
be said to be \textit{closed}.  A whole tree is said to be closed
if all of its branches are closed. Now, in case we want to verify
the inference of a formula~$\alpha$ from a set of
premises~$\gamma_1$, $\gamma_2$, \ldots, $\gamma_n$, using such
2-signed tableau rules for classical logic, what we do is to try
and find a closed tableau tree starting from the linear sequence
of labelled nodes $T{:}\gamma_1$, $T{:}\gamma_2$, \ldots,
$T{:}\gamma_n$, $F{:}\alpha$.

The above tableau system for classical logic respects an obvious
\textit{subformula principle\/} according to which each of the
daughters of a non-closure rule are proper subformulas of some of
the rule heads, disregarding the corresponding labels.  It is easy
to see that the following canonical
\textit{complexity measure\/} decreases with rule application:
\begin{equation}\label{CompMsrCL}
\begin{tabular}{ll}
($\ell1$) & $\ell(p)=0$, where $p$ is an atom\\
($\ell2$) & $\ell(\neg\varphi_1)=\ell(\varphi_1)+1$\\
($\ell3$) & $\ell(\varphi_2\to\varphi_3)=\ell(\varphi_2)+\ell(\varphi_3)+1$%
\end{tabular}
\end{equation}
Obviously, the closure rule is the only rule applicable to nodes
with complexity zero.  We say that a proof system is
\textit{analytical\/} if it only allows you to apply a rule when
its daughters have smaller complexity than at least one of the
corresponding heads.  In other words, an analytical proof system
is one to which a convenient \textit{proof strategy\/} has been conveniently
associated in such a way that complexity always decreases with
rule application. This is obviously the case, without
restriction, for the above collection of rules for classical logic,
applied in any particular order.

Analyticity guarantees \textit{termination\/} of a proof
procedure, as soon as the application of rules has a completely
deterministic result, and becomes otherwise redundant.
We say that a tableau tree is terminated when:
(T1)~all of its branches are closed; (T2)~there are open branches
and no further rule is applicable without introducing redundancies.
In case~(T1) we may say the
the initial inference has been successfully verified;  in
case~(T2), the open branches allow us to extract all the
counter-models to the initial inference.

\section{Many-Valued Logics}

Many-valued logics deviate from classical logic in allowing larger
classes of truth-values, the so-called \textit{designated\/} and
\textit{undesignated\/} values, to represent, respectively,
`degrees of truth' and `degrees of falsity'.  The rest remains
pretty much the same, from the semantical point of view, so that
for each assignment of truth-values to the atoms of a given
$m$-ary formula~$\varphi$ there is a unique way of extending that
into an interpretation~$\widetilde{\varphi}$ of that formula as an
$m$-ary operator over the extended algebra of truth-values.

An algorithm for obtaining analytic 2-signed tableau systems for
finite-valued logics was described in~\cite{ccal:mar:09a}, and we
will illustrate it in what follows, for the instructive case of \L
ukasiewicz's four-valued logic~\L$_4$. This logic has~1 as its
only designated value and~$\frac{2}{3}$, $\frac{1}{3}$ and~$0$ as
its undesignated values.  Its connectives~$\neg$ and~$\to$ are
interpreted as operators over $\mathcal{V}=\{1,\frac{2}{3},\frac{1}{3},0\}$
by way of the following definitions and their corresponding
truth-tables:
\begin{equation}\label{tableL4}
\begin{tabular}{ll}
(\L$_4$$\neg$) & $\widetilde{\neg}v=1-v$\\
(\L$_4$$\to$)  & $v_1\widetilde{\to}v_2=\textsf{Min}(1,1-v_1+v_2)$\\
\end{tabular}
\end{equation}
Now, to produce a classic-like 2-signed tableau system for~\L$_4$
the idea is to associate, in terms of the signs~$T$ and~$F$,
to each truth-value of this logic a unique \textit{binary print\/}
that distinguishes this truth-value from any other truth-value.
Given a collection of truth-values~$\mathcal{V}$, its
characteristic function $t:\mathcal{V}\to\{T,F\}$ is a mapping
that associates~$T$ to designated values and~$F$ to undesignated
values.  Binary prints are sequences of unary formulas, called
\textit{separating formulas}, that use the latter characteristic
functions to distinguish in between truth-values.  In the case
of~\L$_4$, the following choice of separating formulas can be seen to do
the job: $\theta_1(\varphi)=\neg \varphi$ and
$\theta_2(\varphi)=\neg\neg(\varphi\to\neg\varphi)$.  Consider
indeed the table:%
{\small
\begin{equation}\label{binaryPrintsL4}
\begin{array}{|c|c||c|c||c|c|}\hline
                    \ v \ &  t(v) &  \widetilde\theta_1(v) & t(\widetilde\theta_1(v)) & \widetilde\theta_2(v) & t(\widetilde\theta_2(v)) \\ \hline\hline
                    \ 0 \ &         F &                                 1 &                 T &                     1 & T \\ \hline
  \ \frac{1}{3} \ &       F &           \frac{2}{3} &               F &                     1 & T \\ \hline
  \ \frac{2}{3} \ &         F &             \frac{1}{3} &               F & \frac{2}{3} & F \\ \hline
                    \ 1 \ &         T &                                 0 &                 F &                     0 & F\\ \hline
\end{array}
\end{equation}
}%
Notice how each truth-value~$v$ is associated to a unique triple
$\left\langle
t(v),t(\widetilde\theta_1(v)),t(\widetilde\theta_2(v))\right\rangle$.

All rules of the corresponding tableau system will have labelled
binary prints as branches.  For example, the rules corresponding
to (\L$_4$$\neg$) are:%
{\small
\begin{equation}\label{L4negrlz}
\begin{tabular}{cc}
\noindent \Tree
    [.{ $F{:}\neg\alpha$ }
      [.{ $F{:}\alpha\hspace{3mm} $\\$ F{:}\theta_1(\alpha)\hspace{3mm}$\\ $T{:}\theta_2(\alpha)\hspace{3mm}$ }
      ]
      [.{ $F{:}\alpha $\\$ F{:}\theta_1(\alpha)$\\ $F{:}\theta_2(\alpha)$ }
      ]
      [.{ $\hspace{3mm}T{:}\alpha $\\$ \hspace{3mm}F{:}\theta_1(\alpha)$\\ $\hspace{3mm}F{:}\theta_2(\alpha)$ }
      ]
    ]
    \hspace{1cm}
&
    \hspace{1cm}
\Tree
    [.{ $T{:}\neg\alpha$ }
      [.{ $F{:}\alpha $\\$ T{:}\theta_1(\alpha)$\\ $T{:}\theta_2(\alpha)$ }
      ]
    ]
    \smallskip
\end{tabular}
\end{equation}
}%
An additional set of rules, with heads of the form
$S{:}\theta_1(\varphi)$ and $S{:}\theta_2(\varphi)$, with
$\varphi=\neg\alpha$ and $\varphi=\alpha\to\beta$, and
$S\in\{T,F\}$, is needed to guarantee soundness and completeness
of the 2-signed tableau system with respect to the initial
finite-valued truth-tabular characterization of the current target logic,
\L$_4$. Here are, by way of an illustration, the rules for
$T{:}\theta_{2}(\alpha\to\beta)$ and $T{:}\theta_{1}(\neg\alpha)$:%
{\small
\begin{equation}\label{L4trlz}
\begin{tabular}{cc}
\Tree
    [.{ $T{:}\theta_{2}(\alpha\to\beta)$ }
      [.{ $F{:}\alpha\hspace{3mm} $\\$ F{:}\theta_1(\alpha)\hspace{3mm}$\\ $F{:}\theta_2(\alpha)\hspace{3mm}$\\$ F{:}\beta\hspace{3mm}$\\$ T{:}\theta_1(\beta)\hspace{3mm}$\\$ T{:}\theta_2(\beta)\hspace{3mm}$ }
      ]
      [.{ $T{:}\alpha $\\$ F{:}\theta_1(\alpha)$\\ $F{:}\theta_2(\alpha)$\\$ F{:}\beta$\\$ T{:}\theta_1(\beta)$\\$ T{:}\theta_2(\beta)$ }
      ]
      [.{ $\hspace{3mm}T{:}\alpha $\\$ \hspace{3mm}F{:}\theta_1(\alpha)$\\ $\hspace{3mm}F{:}\theta_2(\alpha)$\\$\hspace{3mm}F{:}\beta$\\$\hspace{3mm} F{:}\theta_1(\beta)$\\$\hspace{3mm} T{:}\theta_2(\beta)$ }
      ]
    ]
    \smallskip
    \hspace{1cm}
&
    \hspace{1cm}
\Tree
    [.{ $T{:}\theta_{1}(\neg\alpha)$ }
      [.{ $T{:}\alpha $\\$ F{:}\theta_1(\alpha)$\\ $F{:}\theta_2(\alpha)$ }
      ]
    ]
    \smallskip
\end{tabular}
\end{equation}
}%

Finally, the set of closure rules contains not only the classical
rule~\eqref{ClRl1}, but also all other combinations of labelled
binary prints that do \textit{not\/} correspond to possible
valuations, according to the truth-tables of~\L$_4$.  In the case
of this logic, the extra closure rules will then be:%
{\small
\begin{equation}\label{L4Clrlz}
\begin{tabular}{cccc}
\Tree
    [.{ {$F{:}\alpha $}\\$ T{:}\theta_1(\alpha)$\\ $F{:}\theta_2(\alpha)$ }
      [.{ $\divideontimes$}
      ]
    ]
\hspace{1cm} & \Tree
    [.{ $T{:}\alpha $\\ {$F{:}\theta_1(\alpha)$}\\ $T{:}\theta_2(\alpha)$ }
      [.{ $\divideontimes$}
      ]
    ]
\hspace{1cm} & \Tree
    [.{ {$T{:}\alpha $}\\$ T{:}\theta_1(\alpha)$\\ {$F{:}\theta_2(\alpha)$} }
      [.{ $\divideontimes$}
      ]
    ]
\hspace{1cm} & \Tree
    [.{ $T{:}\alpha $\\ {$T{:}\theta_1(\alpha)$}\\ {$T{:}\theta_2(\alpha)$} }
      [.{ $\divideontimes$}
      ]
    ]
\end{tabular}
\end{equation}
}%
A closer look at the above four closure rules will reveal, for
instance, that the second and fourth rules, from left to right,
only differ in signs for $\theta_{1}(\alpha)$.  Clearly, however,
$T{:}\theta_{1}(\alpha)$ and $F{:}\theta_{1}(\alpha)$ are the
only two possible ways of labelling the
formula~$\theta_{1}(\alpha)$.  Accordingly, those
two rules should give origin to a simpler rule:%
{\small
\begin{equation}\label{L4ClRl}
\Tree
    [.{ $T{:}\alpha$ \\ $T{:}\theta_{2}(\alpha)$}
      [.{ $*$}
      ]
    ]
\end{equation}
}%
A similar approach can in fact be used to simplify other rules of
the system, reducing the number of resulting branches and formulas
(cf.~\cite{mar:men:tfaae4fvl}).  Using that idea, for instance,
the three branches of the rules $[F{:}\neg]$ and
$[T{:}\theta_2\rightarrow]$, in the left halves
of~(\ref{L4negrlz}) and~(\ref{L4trlz}), could be simplified into
just two branches, each with one node less.

Analyticity for the above system is ensured by enforcing a
particular proof strategy that regulates rule applications based on an
adequate non-canonical measure of complexity.  To implement that
strategy, a convenient first step would be to precede
definition~(\ref{CompMsrCL}) by a further clause:
\begin{equation}\label{CompMsrL4}
\begin{tabular}{ll}
($\ell0$) & $\ell(\theta(\varphi))=\ell(\varphi)$, for every separating formula $\theta$\\
\end{tabular}
\end{equation}
Observe how now different clauses of the upgraded definition of
complexity may potentially apply to the same formula~$\varphi$,
if we look at it as a $\theta$-formula or not.  Notice moreover
that the new complexity measure is still well-defined as a
function, once it is read from ($\ell0$) to ($\ell3$), in this
order.  On the other hand, even if we identify a given formula as
a $\theta$-formula, there might be, for instance,
formulas~$\varphi_1$ and~$\varphi_2$ and separating
formulas~$\theta_1$ and~$\theta_2$ such that
$\theta_1(\varphi_1)=\varphi=\theta_2(\varphi_2)$.  In that case,
the rule to be applied should be the one that decreases the
complexity the most, and this `minimality requirement' should also be
conveniently internalized in the above definition of the
complexity measure (check the details in~\cite{ccal:mar:09a}).
For example, the signed formula
$T{:}\neg\neg((\alpha\to\beta)\to\neg(\alpha\to\beta))$ might
equally well be read as an instance of
$T{:}\theta_{1}(\neg((\alpha\to\beta)\to\neg(\alpha\to\beta)))$ or
as an instance of $T{:}\theta_{2}(\alpha\to\beta)$.  The three
choices of reading would result in three different extensions of
a tableau branch having the initial signed formula as one of its
nodes. The first two choices are, according to the right halves
of~(\ref{L4negrlz}) and~(\ref{L4trlz}):
\begin{center}
\begin{tabular}{cc}
Rule $[T{:}\neg]$ is applied: \hspace{1cm} & \hspace{1cm} Rule $[T{:}\theta_1\neg]$ is applied:\smallskip\\
\Tree
    [.{ $T{:}\neg\neg((\alpha\to\beta)\to\neg(\alpha\to\beta))$ }
      [.{ $F{:}\neg((\alpha\to\beta)\to\neg(\alpha\to\beta)) $\\$ T{:}\theta_1(\neg((\alpha\to\beta)\to\neg(\alpha\to\beta)))$\\ $T{:}\theta_2(\neg((\alpha\to\beta)\to\neg(\alpha\to\beta)))$ }
      ]
    ]
    \smallskip
    \hspace{1cm}
&
    \hspace{1cm}
\Tree
    [.{ $T{:}\theta_{1}(\neg((\alpha\to\beta)\to\neg(\alpha\to\beta)))$ }
      [.{ $T{:}((\alpha\to\beta)\to\neg(\alpha\to\beta)) $\\$ F{:}\theta_1(((\alpha\to\beta)\to\neg(\alpha\to\beta)))$\\ $F{:}\theta_2(((\alpha\to\beta)\to\neg(\alpha\to\beta)))$ }
      ]
    ]
    \smallskip
\end{tabular}
\end{center}
The third choice corresponds exactly to the rule pictured at the left
half of~(\ref{L4trlz}).  Clearly, it is in this last and more `concrete'
choice that the rule application results in less complex formulas.
Our tableau strategy should take that into consideration.  To
guarantee in fact that the new complexity measure given
in~(\ref{CompMsrCL}) and~(\ref{CompMsrL4}) continues to be
well-defined as a complexity \textit{function}, one also has to
guarantee that~(\ref{CompMsrL4}) chooses, for a non-atomic
formula~$\varphi$, the separating formula~$\theta$ that results in
`minimally' complex output branches, when the corresponding rule
is applied.  Details of this can be found in~\cite{ccal:mar:09a}
and~\cite{ccal:mar:09b}.  The final tableau strategy of choice is
then to be strictly based on such upgraded complexity measure, in
order to guarantee analyticity.

Just to illustrate the fundamental relevance of such strategy, if
one did not strictly follow it in the above example, one could have opted for
the first choice of reading, that of rule $[T{:}\neg]$,
and then it could be observed that from the
sequence of three resulting daughters, the second would be just the
head of the rule reiterated, and the third would be the more
complex formula
$T{:}\neg\neg((\neg((\alpha\to\beta)\to\neg(\alpha\to\beta)))\to\neg(\neg((\alpha\to\beta)\to\neg(\alpha\to\beta))))$.
The tableau building procedure, in such a situation, would not
necessarily be terminating.

\section{Tactics}

Our axiom extraction program takes as input the definition of a
many-valued logic and generates a file with a theory ready to use
with \texttt{Isabelle}. The theory includes the set of all
tableau rules for the object logic. In addition, taking advantage
of the analytical character of the system defined by the new
algorithm, rewrite rules and tactics for automated theorem proving
are constructed.

In the output file for the logic \L$_4$, the rules for
$\textit{F}{:}\neg\alpha$, $\textit{T}{:}\neg\alpha$,
$\textit{T}{:}\theta_{2}(\alpha\to\beta)$ and
$\textit{T}{:}\theta_{1}(\neg\alpha)$ exhibited at the previous section
are represented in \texttt{Isabelle}'s syntax\footnote{The syntax
employed here is that of \texttt{Isabelle} 2005, and the assisted
proofs are done in the command line interface.} by:

{\footnotesize
\begin{verbatim}
FNeg:    "[| [ $H, F:A0, F:t1(A0), T:t2(A0), $G ] ;
             [ $H, F:A0, F:t1(A0), F:t2(A0), $G ] ;
             [ $H, T:A0, F:t1(A0), F:t2(A0), $G ] |]
               ==> [ $H, F:~(A0), $G ]"

TNeg:    "[| [ $H, F:A0, T:t1(A0), T:t2(A0), $G ] |]
               ==> [ $H, T:~(A0), $G ]"

Tt1Neg:  "[| [ $H, T:A0, F:t1(A0), F:t2(A0), $G ] |]
               ==> [ $H, T:t1(~(A0)), $G ]"

Tt2Imp:  "[| [ $H, F:A0, F:t1(A0), F:t2(A0), F:A1, T:t1(A1), T:t2(A1), $G ] ;
             [ $H, T:A0, F:t1(A0), F:t2(A0), F:A1, T:t1(A1), T:t2(A1), $G ] ;
             [ $H, T:A0, F:t1(A0), F:t2(A0), F:A1, F:t1(A1), T:t2(A1), $G ] |]
               ==> [ $H, T:t2(A0 --> A1), $G ]"
\end{verbatim}}%

\noindent In the above higher-order sequent-style syntax,
the symbol~\texttt{\$} marks a context, and the meta-implication
\texttt{==>} separates the branch representing the current goal
at the right from its subgoals at the left.  A closure rule such
as the first one from~(\ref{L4Clrlz}), is represented as an axiom
of the form:

{\footnotesize
\begin{verbatim}
CR1:    "[ $C1, F:A, $C2, T:t1(A), $C3, F:t2(A), $C4]"
\end{verbatim}}%

We further add to the theory some convenient rewrite rules to
allow the system to recognize given formulas as instances of
separating formulas whenever possible. Only the outermost
formulas may be instantiated as $\theta$-formulas, as this
rewrite is intended to be followed by a rule application, and
there are no rules for formulas with nested $\theta$s.

{\footnotesize
\begin{verbatim}
t1_def:    "S:~A0          == S:t1(A0)"
t2_def:    "S:~~(A0-->~A0) == S:t2(A0)"
\end{verbatim}}%

Again, to guarantee termination of proofs we must follow a convenient order
of instantiation, starting with the rewrite rule that reduce the
most the complexity of the formula, namely the one that
takes~$\theta_2$ into consideration. A tactic for ordered
instantiation, in the case of~\L$_4$, may be defined by:

{\footnotesize
\begin{verbatim}
val auto_rw = (rewrite_goals_tac [t2_def]) THEN
              (rewrite_goals_tac [t1_def]);
\end{verbatim}}%

\noindent where the command {\tt rewrite\_goals\_tac [t2\_def]}
rewrites all formulas of the subgoal using the definition of {\tt
t2\_def}, and similarly for {\tt t1\_def}.  The tactical {\tt
THEN} makes sure that the second line of the above tactic will be
executed only after the first one, and this strategy will
guarantee the correct order of instantiation in the case where
different $\theta$-rules are applicable, in view of the
minimality requirement mentioned in the previous section,
necessary to guarantee analyticity. Here is an illustration of the
use of~{\tt auto\_rw}:

{\footnotesize
\begin{verbatim}
 1. [F:~~(A-->~A), T:~~(A-->~B)]               (* Current state of proof *)
 2. [T:~~((A-->B)-->~(A-->B)), T:~A, F:~~A]

ML> by auto_rw;                                (* Using the tactic *)
 1. [F:t2(A), T:t1(~(A-->~B))]                 (* New state of proof *)
 2. [T:t2(A-->B), T:t1(A), F:t1(~A)]
\end{verbatim}}%

We may now use again the native \texttt{Isabelle}'s tacticals and
construct a tactic for fully automatic theorem proving, by
describing a procedure to exhaustively repeat, for every branch
of the proof tree, the following steps:

\begin{enumerate}
    \item instantiate formulas by rewriting ({\tt auto\_rw}), then
    \item close the branch by applying one of the closure rules or
    \item apply another rule of the system, in some suitable order.
\end{enumerate}

The first step will ensure that the right choice will be made when
multiple rules are applicable to a formula. Next, the tactic tries
to close the branch as soon as possible, to speed-up the process.
If closure is not possible at that stage, the next step will try
to apply another rule of the system, in the most convenient
application order (for instance, postponing branching as much as
possible), and start again. The procedure terminates, due to
the analyticity of the system, and at the end either
\texttt{Isabelle} will deliver a message that says `{\tt No~subgoals!}',
meaning that the proof has been successfully concluded, or else
there will be a list of subgoals
--- open branches --- which are impossible to close and such that
all their formulas have complexity zero, so that no further rule
is applicable.  From those open branches, as usual, counter-models
can be assembled.

Extra details will be at hand to be surveyed by the interested
reader as the full system is made available on-line, in open
source.


 \providecommand{\url}[1]{#1} \newcommand{\noopsort}[1]{}

\end{document}